# High-$T_C$ Superconductivity Originating from Interlayer Coulomb Coupling in Gate-Charged Twisted Bilayer Graphene Moiré Superlattices

**Dale R Harshman** [1] · **Anthony T Fiory** [2]

**Abstract**

Unconventional superconductivity in bilayer graphene has been reported for twist angles θ near the first magic angle and charged electrostatically with holes near half filling of the lower flat bands. A maximum superconducting transition temperature $T_C \approx 1.7$ K was reported for a device with θ = 1.05° at ambient pressure and a maximum $T_C \approx 3.1$ K for a device with θ = 1.27° under 1.33 GPa hydrostatic pressure. A high-$T_C$ model for the superconductivity is proposed herein, where pairing is mediated by Coulomb coupling between charges in the two graphene sheets. The expression derived for the optimal transition temperature, $T_{C0} = k_B^{-1}\Lambda(|n_{opt} - n_0|/2)^{1/2}e^2/\zeta$, is a function of mean bilayer separation distance ζ, measured gated charge areal densities $n_{opt}$ and $n_0$ corresponding to maximum $T_C$ and superconductivity onset, respectively, and the length constant $\Lambda = 0.00747(2)$ Å. Based on existing experimental carrier densities and theoretical estimates for ζ, $T_{C0} = 1.94(4)$ K is calculated for the θ = 1.05° ambient-pressure device and $T_{C0} = 3.02(3)$ K for the θ = 1.27° pressurized device. Experimental mean-field transition temperatures $T_C^{mf} = 1.83(5)$ K and $T_C^{mf} = 2.86(5)$ K are determined by fitting superconducting fluctuation theory to resistance transition data for the ambient-pressure and pressurized devices, respectively; the theoretical results for $T_{C0}$ are in remarkable agreement with these experimental values. Corresponding Berezinskii-Kosterlitz-Thouless temperatures $T_{BKT}$ of 0.96(3) K and 2.2(2) K are also determined and interpreted.

Dale R. Harshman
drh@physikon.net

Anthony T. Fiory
fiory@alum.mit.edu

[1] Department of Physics, The College of William and Mary, Williamsburg, VA 23187, USA
[2] Department of Physics, New Jersey Institute of Technology, Newark, NJ 07102, USA

# 1 Introduction

Gate charged in field-effect transistor devices, twisted bilayer graphene (TBG) exhibits unconventional superconductivity at "magic" twist angles θ, with the electronic phases of Moiré superlattices determined by θ, gated charge areal density $n$ [1, 2], and pressure-tunable interlayer separation [2, 3]. Noteworthy characteristic features are dome shaped regions in temperature-vs-$n$ diagrams, similar to those observed in superconducting high-$T_C$ cuprates [1, 4], strong-coupled superconductive pairing which apparently cannot be explained by weak electron-phonon coupling [1], and gated scanning tunneling spectroscopy (STS) suggesting electronic pairing [4]. Superconductivity, occurring for hole doping of TBG for $n$ near half flat-band filling, is reported for a θ = 1.05° device at ambient pressure with $T_C$ ≈ 1.7 K (designated "device M2" in [1]) and for a θ = 1.27° device with $T_C$ ≈ 3.1 K ("device D2" at an applied hydrostatic pressure of 1.33 GPa in [2]). Most importantly, the behavior reported in [2] confirms that the fundamental requirement for superconductivity in TBG is formation of a Moiré superlattice with flat electron bands. Current theoretical viewpoints on the superconductivity consider involvement of electron-phonon interactions [5–18] and novel electronic-based alternatives without phonon involvement [19–45]. Works considering phonon [5–11] and electronic [19] interactions have predicted $T_C$ values ranging from 0.04 to 10 K. The tight-binding description of relaxed TBG near magic θ and half-filling of holes in [7] shows two types of Fermi surface sheets, located near the Moiré Brillouin zone boundary and zone center. Electrostatic effects in electron-electron interactions within low-energy bands are shown to be favorable for superconductivity in only one type of Fermi pocket [28].

As demonstrated by the semiconducting interlayer transport reported for TBG, the twist rotation decouples electronic coherence between the two graphene layers, yielding anisotropic transport similar to the cuprates [46]. High resistivity transverse to the layers [47] along with the suppressed interlayer tunneling in TBG shows that carriers occupying the flat bands are effectively localized into separate layers. Observations of interlayer excitons by photo-luminescence [48] provide further evidence of strong correlation between the carriers in the separate layers.

The electronic features of TBG reported in [1, 4, 46-48] indicate the importance of Coulomb interactions in the superconductivity mechanism. Properties such as bilayer structure and interlayer electronic transport decoupling are further suggestive of a Coulomb-based pairing interaction occurring between the two layers. Such a mechanism was introduced for determining the optimal transition temperature $T_{C0}$ of high-$T_C$ cuprates and other superconductors in [49], with further developments most recently reported in [50]. Pairing in this model is ascribed to Coulomb interactions between charges in adjacent reservoirs separated by a distance ζ via the exchange of virtual photons, with $T_{C0}$ scaling with the root of the participating charge density 1/ℓ multiplied by the potential energy $e^2/\zeta$ (where $e$ is the electron charge). In prior work, this approach has been successfully applied to 51 compounds from nine unique superconducting families, ascertained by accurately deriving $T_{C0}$ [49-56]. For TBG devices, the two gate-charged graphene layers, separated by average distance ζ, comprise the interacting reservoir structure. Identifying the participating charge density as the difference between the optimal and superconducting onset in gated charge densities, denoted as $n_{opt}$ and $n_0$, respectively, $T_{C0}$ = 1.94(4) K is calculated for device M2 (ambient pressure) and $T_{C0}$ = 3.02(3) K is found for device D2 (1.33 GPa), illustrating the unique tunability of the TBG system.

For uniform superconducting films, attribution of resistive broadening in the superconducting transition is typically given to fluctuation effects above and below the transition $T_C$. Specifically, in the $T > T_C$ regime, fluctuations in conductivity are modeled using the Aslamasov-Larkin (AL) theory [57], whereas the Berezinskii-Kosterlitz-Thouless (BKT) formalism [58-60] describes superconducting phase fluctuations below $T_C$. Both the AL theory and the Halperin-Nelson (HN) interpolation formula [61], generalized here to accommodate both clean and dirty limits, are applied in the analysis of resistance data indicating highest $T_C$ in [1] and [2], yielding mean-field superconductive transition temperatures $T_C^{mf}$ of 1.83(5) K and 2.86(5) K, and vortex-pair unbinding temperatures $T_{BKT}$ of 0.96(3) K and 2.2(2) K, for devices M2 and D2, respectively.

Application of the interlayer Coulomb pairing model and prediction of $T_{C0}$ is presented in Section 2. Analysis of the longitudinal resistance from the perspective of fluctuation effects above and below $T_C$, arguments for identifying the optimal interaction charge densities and ζ values for devices M2 and D2 are provided in Section 3. Section 4 discusses the theoretical upper limit on $T_C$, effects of inhomogeneities in devices M2 and D2, estimations of the sheet magnetic penetration depth and effective mass in device D2, and statistical analysis of the model in general. Conclusions are drawn in Section 5.

## 2 Interlayer Coulomb Pairing Model: Calculating $T_{C0}$

Superconductive coupling in layered high-$T_C$ superconductors is understood herein as arising from Coulomb interactions between charges in spatially separated charge reservoirs. Originally developed to describe indirect Coulomb pairing interactions between charges in adjacent extended layers [49], this unique approach is also found to be applicable to certain three-dimensional (3D) extended lattices such as $Cs_3C_{60}$ [51] and $H_3S$ [50], where the latter also comprises electronically equivalent charge reservoirs like TBG. This theory predicts $T_{C0}$ for optimal superconductors, which are identified experimentally by maxima in transition temperatures governed by controlling doping, structure, and externally applied pressure. Calculations of $T_{C0}$ have been previously validated against experiment with statistical accuracy of ±1.30 K or ±4% in $T_{C0}$ for 51 different superconductors from nine superconducting families, consisting of the aforecited 3D compounds [50, 51], layered cuprates, ruthenates, rutheno-cuprates, iron pnictides, BEDT-based [bis(ethylenedithio)tetrathiafulvalene] organics [49, 52, 53], iron chalcogenides [54], and intercalated group-4-metal nitride-chlorides [55, 56], with measured superconducting transitions ranging from ~7 to 200 K.

The unconventional superconductivity present in TBG devices clearly arises from electronic correlations in a manner consistent with high-$T_C$ materials [1]. The high-$T_C$ nature of TBG devices is further supported circumstantially given that the two requisite physically separate charge reservoirs, denoted as types I and II in [49] as comprising the interaction structure of the model, are present in the form of the two individual graphene sheets which are separated by mean distance ζ. Because they are essentially identical, the graphene layers must each function as both reservoir types, with pairing carriers and mediating charges coexisting in each layer, much like as was found for the two interlaced sublattices forming the $Im\bar{3}m$ structural phase of $H_3S$ [50]. The coexistence of superconducting and non-superconducting pockets in TBG devices as surmised in [22, 28], together with strong interlayer electron-hole correlations [48], is consistent with the two reservoir types required.

The optimal transition temperature $T_{C0}$ of a homogeneous high-$T_C$ superconductor is given by the algebraic formula [49],

$$T_{C0} = k_B^{-1} (\Lambda/\ell) e^2/\zeta, \qquad (1)$$

where ζ is the perpendicular distance separating the two interacting charge reservoirs, i.e., the two graphene sheets, the length $\ell = (\sigma\eta/A)^{-1/2}$ is the linear interaction charge spacing defined by the participating charge σ per unit area $A$, multiplied by the number of mediating (type II) layers η (equal to unity for TBG devices), and the length constant $\Lambda = 0.00747(2)$ Å $= [1.933(6)]\lambdabar_C$ that is previously determined experimentally in [49] and expressed as nearly twice the reduced electron Compton wavelength $\lambdabar_C$.

The participating charge is in general defined by the expression $\sigma = \gamma[v|x_{opt} - x_0|]$, where [$x_{opt}$] is the optimal doping charge required for superconductivity at highest measured $T_C$, [$x_0$] is the threshold for the onset of superconductivity, γ is the sharing factor and $v$ is the dopant valence. Likening gating to doping, the factor γ follows the charge allocation rule 1b in [49] for sharing the doping charge with the two charge reservoirs, which specifies γ = 1/2, and electronic charge with $v \equiv 1$ is bounded by $0 \le [x] \le 4$. In reference to the TBG superlattice, σ is the participating charge per Moiré superlattice unit cell of area $A =$ $3^{1/2}a^2/[8\sin^2(\theta/2)]$ in the absence of strain,[1] where $a$ is the graphene lattice constant (nominally 2.46 Å) [28]. In terms of gated charge areal density $n$, the net doping charge areal density is $[|x_{opt} - x_0|]/A = |n_{opt} - n_0|$, where $n = n_{opt}$ yields highest measured $T_C$ and $n = n_0$ is at the boundary between insulating and superconducting states (see Section 3.2.1). One thus has $\ell = (|n_{opt} - n_0|/2)^{-1/2}$. Incor-porating these definitions specific to TBG devices, Eq. (1) is written,

$$T_{C0} = k_B^{-1}\Lambda(|n_{opt} - n_0|/2)^{1/2}e^2/\zeta. \qquad (2)$$

**Table 1.** Results for TBG devices M2 from [1] and D2 (at 1.33 GPa) from [2] for the interlayer Coulomb pairing model calculation of Eq. (2). Parameters are mean interlayer spacing $\zeta$, optimal gated charge density $n_{opt}$, superconductive onset gated charge density $n_0$, sharing factor $\gamma$, mean spacing between participating charges $\ell$, and calculated optimal transition temperature $T_{C0}$, as defined in the text.

| TBG device | Interlayer Coulomb pairing model parameters | | | | | |
| --- | --- | --- | --- | --- | --- | --- |
| | $\zeta$ (Å) | $n_{opt}$ ($10^{12}$ cm$^{-2}$) | $n_0$ ($10^{12}$ cm$^{-2}$) | $\gamma$ | $\ell$ (Å) | $T_{C0}$ (K) |
| M2 | 3.50(1) | −1.44(2) | −2.03(1) | 1/2 | 184(4) | 1.94(4) |
| D2 | 3.42(1) | −2.11(2) | −3.47(2) | 1/2 | 121(1) | 3.02(3) |

In relation to the theoretical definition of the electron areal density $n_S = 4/A$ in the charge-neutral superlattice, experiment finds $n_{opt}$ to be near $-n_S/2$ and $n_0$ approaching $-n_S$, as discussed below. However, the parameters $n_S$, $A$, and $\theta$ for a given device need not be specified for calculating either $\ell$ or $T_{C0}$ from Eq. (2).

## 2.1 Electronic and Structural Parameters for Devices M2 and D2

The participating charge density $|n_{opt} - n_0|/2$ can be extracted from measurements of the phase diagram. For TBG device M2, the optimal and onset carrier densities, $n_{opt}$ −1.44(2) × $10^{12}$ cm$^{-2}$ and $n_0$ = −2.03(1) × $10^{12}$ cm$^{-2}$, respectively, are extracted from the data for device M2 in [1] as described in Section 3.2.1, such that $\ell = 184(4)$ Å.

---
[1] Strains of 0.6 to 0.7% in uncapped magic-angle TBG structures were determined from scanning tunneling microscopy [4].

Similarly, $n_{opt}$ and $n_0$ from data in [2] and described in Section 3.2.2 are −2.11(2) × $10^{12}$ cm$^{-2}$ and −3.47(2) × $10^{12}$ cm$^{-2}$, respectively, giving $\ell = 121(1)$ Å.

Unfortunately, bilayer separations in magic angle TBG are somewhat difficult to determine experimentally from scanning microscopies based on atomic force (AFM) [62] or electron tunneling (STM) [4, 63-66]. Consequently, theoretical results [7, 67-69] and insights from available data are used to establish $\zeta$ following the methodology described in Section 3.2.3. An ambient-pressure value of $\zeta = 3.50(1)$ Å is obtained for device M2, and from considerations of TBG under uniaxial pressure [3, 70] with an elastic anisotropy approximating graphite [71], $\zeta = 3.42(1)$ Å is found for device D2 at a hydrostatic pressure of 1.33 GPa.

Using these numbers, Eq. (2) thus gives $T_{C0} = 1.94(4)$ K and 3.02(3) K for TBG devices M2 and D2 (at 1.33 GPa), respectively. All relevant parameters are listed in Table 1.

## 3 Methodology

Below are presented detailed analyses of resistance data for TBG devices M2 [1] and D2 [2], from which are extracted corresponding values of the mean-field transition temperature $T_C^{mf}$ and the Berezinskii-Kosterlitz-Thouless temperature $T_{BKT}$ at optimal doping. Evaluations of the participating charge components $n_{opt}$ and $n_0$, and $\zeta$ for these devices are then presented.

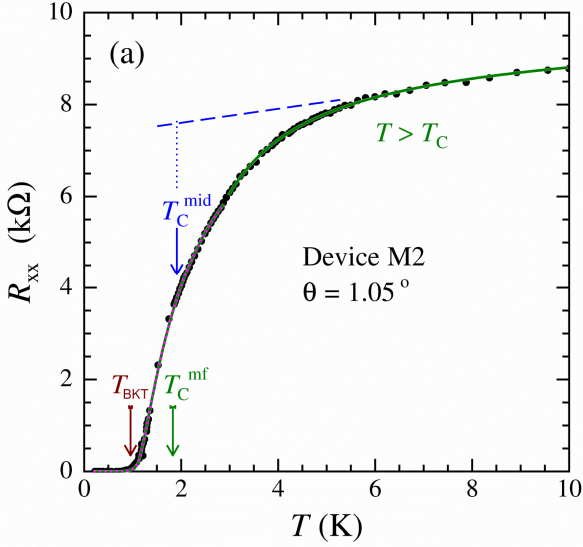

**Fig. 1 (a)** Resistance $R_{xx}$ vs. temperature $T$ for TBG device M2 from [1] (filled circles) overlaid with fitted functions for fluctuation conductance (solid curve) and fluctuation resistance (dotted curve). Arrows mark fits for $T_C^{mf}$ and $T_{BKT}$ (bars denote uncertainties). Midpoint $T_C^{mid}$ is determined from dashed line drawn tangent to data for $T > 6$ K.

## 3.1 Analysis of Longitudinal Resistance Data

Longitudinal resistance data measured as a function of temperature $R_{xx}(T)$ are presented in [1] for two gate-charged twisted bilayer graphene (TBG) devices, denoted M1, $\theta = 1.16°$ and M2, $\theta = 1.05°$. The superconductive transition and Berezinskii-Kosterlitz-Thouless temperatures for device M2 are stated in [1] to be $T_C \approx 1.7$ K and $T_{BKT} = 1.0$ K, derived from 50% of normal-state resistance and non-linearity in current-vs-voltage curves, respectively; for device M1, $T_C$ is approximately 0.5 K. Gated charge density $n$ determines the phases described in [1], which include two superconducting domes straddling a correlated "Mott" insulator for device M1 and merged superconducting domes for device M2. Given that device M2 becomes fully superconducting, as opposed to the weakly superconducting device M1 [1], the focus of this analysis is on the data reported for device M2.

Broadening and rounding of $R_{xx}(T)$ extending to temperatures well above the stated values of $T_C$ are attributed herein to superconductive conductivity fluctuations in the normal state [57],

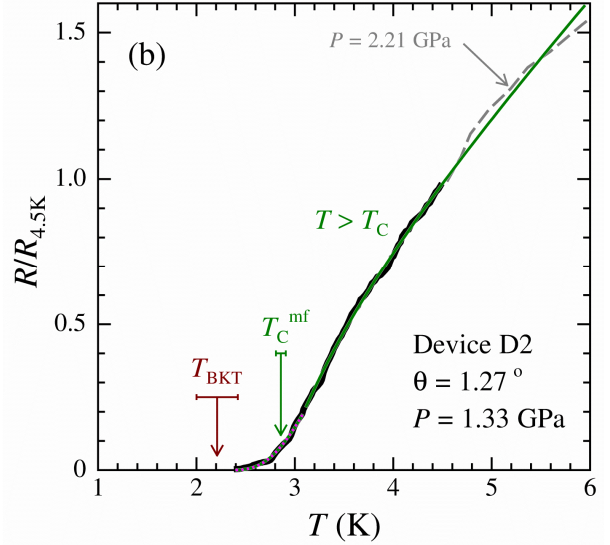

**Fig. 1 (b)** Normalized $R/R_{4.5K}$ vs. $T$ for pressurized (1.33 GPa) TBG device D2 from [2] (thick solid curve) overlaid with fitted functions and marked as in panel (a). Dashed curve represents data at 2.21 GPa and $T > 4.5$ K [2].

phase fluctuations in the superconducting state below $T_C$ [58-61], and materials inhomogeneities as discussed in [1]. As shown below, modeling the fluctuation behavior of $R_{xx}(T)$ in the manner as described yields quantitative evaluations of $T_C$ and $T_{BKT}$.

Figure 1(a) shows the data for $R_{xx}(T)$ of device M2 transcribed from Fig. 1b of [1].[2] The dashed line segment, drawn tangentially to the data points for $T > 6$ K, represents the constructed linear trend of the normal-state resistance. An arrow marks the transition midpoint $T_C^{mid} = 1.91(3)$ K where $R_{xx}(T_C^{mid})$ is 50% of the dashed line value; $T_C^{mid}$ is somewhat higher than the similarly measured $T_C \approx 1.7$ K in [1]. A residual sheet resistance of $r_{0\square} = r_0/(L/W) = 7.35$ k$\Omega/1.5 = 4.9$ k$\Omega/\square$ is indicated from the $T \to 0$ extrapolation of the dashed line in Fig. 1 and the length-to-width aspect ratio $L/W \approx 1.5$ from the Hall bar pattern depicted in the Fig. 1b inset of [1]. This provides an indication of the product of Fermi wave vector and mean free path as $k_Fl = (h/e^2)/r_{0_\square} \approx 5.3$, signifying fairly pure 2D metallic conductance in the gated TBG material. Also, $r_{0_\square} < h/4e^2 = 6.45$ kΩ/□ places device M2 within the upper limit for sustaining superconductivity [72]. The value $k_F = (A_{FS}/\pi)^{1/2} \approx 0.19$ nm$^{-1}$ is calculated from Fermi sheet area $A_{FS} = (2\pi)^2 n_e/g = 0.11$ nm$^{-2}$, electron density $n_e = [x]/A = n_{opt} - (-n_S) = 1.12(2) \times 10^{12}$ cm$^{-2}$, and degeneracy $g = 4$ for spins and valleys [73]. Comparison of the resulting $l \approx 28$ nm to $\xi_{GL} \approx 52$ nm reported for device M1 in [1] (and assumed for device M2) suggests a dirty correction factor of $l/\xi_0 \approx 0.46$, where $\xi_0$ is the Pippard coherence distance.

---

[2] Data points are taken at the visible dots and at approximate locations where the dots appear to be obscured. Line segments connecting dots are assumed to be guides to the eye and are ignored.

Figure 1(b) shows the temperature dependence of normalized longitudinal resistance $R/R_{4.5K}$ for device D2 at applied pressure $P = 1.33$ GPa, transcribed from the semi-log plot Fig. 2C of [2] and displayed here on linear scales. A transition temperature of ≈3.1 K was obtained in [2] from the crossover point in linear fits of the low- and high-temperature regions of the resistance curve on a logarithmic scale. Variation of $R/R_{4.5K}$ at 2.21 GPa for $T > 4.5K$ is also provided for comparison (dashed curve, from Fig. 3C in [2]). Fluctuation analysis of the data in Figs. 1(a) and 1(b) are presented in Sec. 3.1.3.

### 3.1.1 Theoretical Model for $R_{xx}(T)$ for $T > T_C$

The mean field transition temperature $T_C^{mf}$ is determined from fitting $R_{xx}(T)$ using the asymptotic form of the Aslamasov-Larkin (AL) theory [57, 74] for the superconductive fluctuation conductivity above $T_C$, as expressed by the function,

$$R_{FC}(T) = [R_N^{-1} + S(T/T_C - 1)^{-1}]^{-1}\;; T > T_C\,,\quad(3)$$

where the normal-state resistance $R_N = r_0 + r_1T$ is modeled with linear temperature dependence [1, 73], and the factor $S$ is treated empirically. Equation (3) is taken to represent the AL [57], Maki-Thompson [75, 76] and density of states contributions to the fluctuation conductivity, which are expected to have similar forms [74]. For a clean-limit superconductor, the prefactor in AL theory is $S = (W/L)(e^2/16\hbar)$ [57], while the dirty-limit case has the form $S \propto R_N^{-1}$. The coefficient $S$ is generalized by interpolating the forms for clean and dirty superconductors giving,

$$S = c_1(e^2/16\hbar) + c_2 R_N^{-1}\,,\quad(4)$$

with numerical constants $c_1$ and $c_2$ of order unity.

### 3.1.2 Theoretical Model for $R_{xx}(T)$ for $T \gtrsim T_{BKT}$

The low temperature regime, which includes the onset of finite $R_{xx}$, is assumed to be governed by superconducting phase fluctuations in 2D. Fluctuation resistance is modeled by the interpolation formula derived for superconducting vortex-pair excitations, expressed in terms of the (Berezinskii-Kosterlitz-Thouless) vortex-pair unbinding temperature $T_{BKT}$ as [61],

$$R_{HN}(T) = \{c_3 R_N^{-1}\sinh^2[(b\tau_c/\tau)^{1/2}] + R_N^{-1}\}^{-1}\;;\quad T \geq T_{BKT},\quad(5)$$

where $\tau = (T - T_{BKT})/T_{BKT}$ and $\tau_c = (T_C - T_{BKT})/T_{BKT}$; $c_3$ and $b$ are numerical (non-universal) constants. The clean case is treated by the Bardeen-Stephen model of free flux-flow resistance, written as $R_f = (n_f\phi_0/H_{C2})R_N$ [77], which depends on the density of thermally excited unpaired vortices $n_f = 2\pi C_1\xi_+^{-2}$, the order parameter correlation length $\xi_+ = a\xi_c\exp[(b\tau_c/\tau)^{1/2}]$, the coherence distance $\xi_c$, and $\phi_0/H_{C2} = 2\pi\xi_c^2$; numerical constants $C_1$ and $a$ are given in Eqs. (22a) and (22b) of [61]. This results in Eq. (5) with $c_3 = 2\pi C_1/a^2$, whereas the derivation for the dirty-limit leads to $c_3 = 0.37/b$, as given by Eq. (35) in [61].

Since the TBG devices have finite size $D$, the theoretical singularity in $\xi_+$ leading to $R_{HN} = 0$ at $T_{BKT}$ is removed by the constraint $\xi_+ \leq D$, where the equality $\xi_+ = D$ corresponds to the excitation of free vortices of density $n_f' \approx D^{-2}$ and occurs at a temperature $T'$ determined from Eq. (5). By setting $R_{HN} = 2\pi\xi_c^2 n_f'R_N$, which invokes the free flux flow limit [77], one obtains $T' = (1 + \tau')T_{BKT}$, where $\tau' = b\tau_c \operatorname{arcsinh}^{-2}[(2\pi c_3\xi_c^2 n_f')^{-1/2}]$. Free vortices at $T'(\approx T_{BKT})$ are thermally activated with renormalized vortex-pair interaction energy $U_V \approx 4k_BT_{BKT}\ln(D/\xi_c)$. The device $R_{xx}(T)$ for $T < T'$ is thus treated using thermally activated resistance as,

**Table 2.** Parameters from fitting the resistance transitions for gate-charged TBG devices M2 [1] and D2 (at 1.33 GPa) [2], $R_{xx}(T)$ and normalized $R/R_{4.5K}(T)$, respectively. Fluctuation conductance in the normal state is fitted for $T > T_X$ from Eqs. (3) and (4); fluctuation resistance in the superconducting state is fitted for $T < T_Y$ from Eqs. (5) and (6). Parameters are defined in the text.

| TBG device | Normal resistance | | Conductance fluctuations ($T > T_X$) | | | | Resistance fluctuations ($T < T_Y$) | | | | |
|---|---|---|---|---|---|---|---|---|---|---|---|
| | $r_0$ | $r_1$ | $T_C^{mf}$ (K) | $c_1$ | $c_2$ | $T_X$ (K) | $T_{BKT}$ (K) | $T'$ (K) | $b$ | $c_3$ | $T_Y$ (K) |
| M2 | 9.27(8) kΩ | 0.021(15) kΩ/K | 1.83(5) | 0.67(11) | 0.25(2) | 3.0(5) | 0.96(3) | 1.11(5) | 1.37(4) | 0.79(3) | 2.9(8) |
| D2 | 0.02(2) [a] | 0.36(1)/K [a] | 2.86(5) | 1.1(5) [b] | 0.36(1) | 3.1(2) | 2.2(2) | 2.4(2) | 10.5(6) | 0.07(1) | 3.1(3) |

[a] normalized to $R_{4.5K}$
[b] $R_{4.5K} = 1.16$ kΩ assumed

$$R'(T) = 2\pi (\xi_c/D)^2 R_N \exp[U_v/k_B T' - U_v/k_B T] \; ; \quad (6)$$
$$T \leq T',$$

which includes temperatures below $T_{BKT}$. For device M2, $D = 1$ μm is taken from the square-root of the 1-μm² device area cited in [1] and $\xi_c = \xi_{GL} = 52$ nm, since $T' \ll T_C$. For device D2, one correspondingly takes $D = 1$ μm and $\xi_{GL} = 38$ nm from [2]. Resistance fluctuations are thus approximated by Eq. (5) for $T \geq T'$ and Eq. (6) for $T \leq T'$.

### 3.1.3 Fitting Criteria and Results

The applicable range of Eq. (3) is constrained by $T > T_C$. Therefore, resistance-vs.-temperature data are modeled by Eqs. (3) and (4) in the region $T > T_X$, where the cut-off temperature $T_X$ is treated as a fitting parameter. Eqs. (5) and (6) apply to the low-temperature regime $T < T_Y$, where the fitted parameter $T_Y > T_C$ marks the upper limit cut-off and dependent variable $T'$ is defined above. The model is fitted to the $R_{xx}(T)$ data for device M2 (Fig. 1(a)) and $R/R_{4.5K}(T)$ data at 1.33 GPa for device D2 (Fig. 1(b), solid curve). In the latter case, $R_{4.5K} \approx 1.16$ kΩ from resistance data at $P = 2.21$ GPa in Fig. 3C in [2] is assumed in deriving the coefficient $c_1$ for the clean-limit term in the fluctuation conductivity. Parameters that best fit the model functions to the data are determined by a non-linear simplex method for chi-square minimization. The chi-square (total variance per degree of freedom) is $2.156 \times 10^{-3}$ kΩ² in fitting $R_{xx}(T)$ for device M2 and $9.56 \times 10^{-5}$ in fitting $R/R_{4.5K}(T)$ for device D2. Results are provided in Table 2, where $T_C^{mf}$ denotes the fitted value for $T_C$; note that $T_X \geq T_Y$. Uncertainties in the least significant digits originate from averages of the +/− parameter deviations corresponding to a doubling of the chi-square.

For $R_{xx}(T)$ of device M2, the fitted function for conductance fluctuations at $T > T_X$ is shown in Fig. 1(a) as the solid curve and labeled $T > T_C$; the function for resistance fluctuations at $T < T_Y$ is denoted by the dotted curve. Arrows indicate fitted $T_C^{mf} = 1.83(5)$ K and $T_{BKT} = 0.96(3)$ K (arrow-tail widths are commensurate with the uncertainties). The fitted $T_C^{mf}$ is close to the midpoint construction for $T_C^{mid}$; the fitted $T_{BKT}$ is close to the 1.0 K extracted from current-voltage curves in [1]. These two temperature points set $\tau_c = 0.91 \pm 0.11$, which is larger than that usually found for vortex-pair unbinding transitions in thin superconductor films (i.e. $\tau_c < 0.5$ is reported in [78, 79]) and is more typical for weak-link arrays (e.g. device dependent $\tau_c$ ranges from ~ 0.8 to 2 in [80]).

Results of fitting $R/R_{4.5K}(T)$ at $P = 1.33$ GPa for device D2 are shown in Fig. 1(b) as the solid curve for conductance fluctuations in the normal state (extrapolated for $T > 4.5$ K) and the dotted curve for resistance fluctuations in the superconducting state. Arrows indicate fitted $T_C^{mf} = 2.86(5)$ K and $T_{BKT} = 2.2(2)$ K; $\tau_c = 0.3 \pm 0.1$ thus indicates consistency with treating this compressed TBG as a thin superconducting film. Uncertainties include estimated systematic uncertainty in $R_N(T)$ arising from the restriction $T \leq 4.5$ K.

While the resistance transition data appear to be well described without including corrections for inhomogeneity in $T_C$, some inhomogeneity does appear to be present [1, 2], which is attributed in [2] and [4] to local strains and variabilities in twist angle, such as found in similarly fabricated TBG structures by STM [4] and transmission electron microscopy [81]. Wrinkles, spatially evolving twists and extreme strain regions are also observed over large areas [4].

### 3.2 Determining $n_{opt}$, $n_0$, and $\zeta$

A parallel between the measured changes in electronic phase as function of gated charge density in TBG devices and similar phase variations with chemical doping in high-$T_C$ cuprate superconductors is noted in [1]. Other commonalities are that comparatively higher transition temperatures tend to be associated with comparatively higher normal-state resistances, which is manifested in gated TBG by comparing different devices or different gated charges for a given device [1, 73], and longitudinal resistance increases linearly with temperature [73]. As previously indicated, the participating charge density per Moiré unit cell is defined as $|n_{opt} - n_0|/2$, with the participating charge defined as $|x_{opt} - x_0|/2 = A/\ell^2$. The methods for determining these values for devices M2 and D2, and associated interaction distances $\zeta$, are provided below.

### 3.2.1 Identification of $n_{opt}$ and $n_0$ for Device M2 (Ambient Pressure)

The phase diagram for device M2 from Fig. 2c of [1], which is similar to that of the cuprates, exhibits a region of superconductivity for gated charge between an insulator phase reflecting underdoping and a non-superconducting metallic phase corresponding to overdoping, in reference to doping level determined by [x]. As determined in [1], $n_{opt} = -1.44(2) \times 10^{12}$ cm$^{-2}$ (uncertainty approximated from darkest blue region of Fig. 2c), which lies below half band filling estimated in the range of $n \approx -(1.25$ to $1.35) \times 10^{12}$ cm$^{-2}$. Following the parallels with the cuprates, the onset of superconductivity is at the underdoping limit, establishing the criterion $n_0 < n_{opt}$. The insulator phase in device M2 occurs at underdoping for $n < -2.03(1) \times 10^{12}$ cm$^{-2}$ (read from Fig. 6a in [1]). Positive magneto-resistance is observed throughout the region between insulator and finite $T_C$, indicating the presence of two-dimensional fluctuation superconductivity that becomes quenched by application of transverse magnetic fields. Taking the fluctuation superconductivity into account fixes the doping onset of superconductivity at $n_0 = -2.03(1) \times 10^{12}$ cm$^{-2}$. The superconducting onset behavior resembles that of underdoped La$_{2-x}$Sr$_x$CuO$_4$, where vestigial superconductivity is in evidence at nominally zero $T_C$ [82].

Measurements of the electron Hall number density $n_H$ (i.e. referenced to negative charge) at $T = 0.4$ K indicate reduced Hall hole densities $-n_H < -n$ in the region $n_0 < n < -n_S/2$, relative to the actual gated charge density; e.g. measurement at $n = n_{opt}$ gives $n_H = -0.49 \times 10^{12}$ cm$^{-2}$ = 0.34 $n_{opt}$ (from extended data Fig. 3 in [1]). On the overdoping side of the superconducting dome, a nearly complete hole density $-n_H \approx -n$ is observed for $n > -n_S/2$.

Band properties as functions of $n$ in the vicinity of $n \sim -n_S/2$ are determined in [1] from normal-state quantum oscillations in the resistance of device M2 in applied magnetic fields. For $-n_S < n < -n_S/2$, anomalous Landau-level structure indicating twofold band degeneracy is observed, whereas single band fourfold degeneracy is evident for $-n_S/2 < n < 0$. From Shubnikov-deHaas oscillations, the band with twofold degeneracy has high effective mass $m^*$ that decreases as $n$ approaches $-n_S/2$ from below, while the band with fourfold degeneracy has low $m^*$ that increases as $n$ approaches $-n_S/2$ from above. Owing to lack of resolved quantum oscillations, these methods are unable to discern band properties near half-filling and at $n = n_{opt}$ in particular.

The resulting participating charge density for calculating $T_{C0}$ is $|n_{opt} - n_0|/2 = 3.0(1) \times 10^{11}$ cm$^{-2}$. Note that the two-dimensional density $n_{2D} = 1.5 \times 10^{11}$ cm$^{-2}$ noted in [1] is a different quantity that is determined therein from optimal doping relative to the half-filling state. Given that $\ell$ = 184(4) Å and $A$ = 156.1 nm$^2$ ($\theta$ = 1.05°), the participating charge $|x_{opt} - x_0|/2 = A/\ell^2 = 0.46(2)$.

### 3.2.2 Identification of $n_{opt}$ and $n_0$ for Device D2 at 1.33 GPa

Optimal doping for device D2 is stated to be roughly at the blue arrow marked in Fig. 2A in [2], which points to $n \approx -2.2 \times 10^{12}$ cm$^{-2}$. To obtain greater precision, $n_{opt}$ for device D2 at 1.33 GPa is extracted from the maxima displayed in the colored resistance contour maps of Figs. 3A and 3B in [2], which exhibit dome-like variations with $n$ in temperature and magnetic field, respectively. Using the gated charge associated with the highest temperature or magnetic field (both being the same) yields $n_{opt} = -2.11(2) \times 10^{12}$ cm$^{-2}$. The underdoped regime appears similar to that of device M2, where Fig. 4A in [2] exhibits an onset of positive magnetoresistance at $n_0 = -3.47(2) \times 10^{12}$ cm$^{-2}$ that is similarly offset as $n_0 > -n_S$. This corresponds to a participating charge $|x_{opt} - x_0|/2 = A/\ell^2$ of 0.73(1), for $\ell$ = 121(1) Å, $A$ = 106.7 nm$^2$ ($\theta$ = 1.27°), and assuming $a$ = 2.46 Å.

Hall effect data for device D2 at $P$ = 1.33 GPa [2] shows electron-like response in the superconducting dome region with $n_H \approx +0.24 \times 10^{12}$ cm$^{-2}$ = $-0.11 n_{opt}$ measured at $n = n_{opt}$. This contrasts with the hole-like $n_H$ generally observed for TBG devices in [1]. At ambient pressure and $P$ = 2.21 GPa (both being non-optimal for superconductivity), the Hall effect measured at the same $n = n_{opt}$ is hole-like, albeit of reduced magnitude, with $n_H = -0.23 \times 10^{12}$ cm$^{-2}$ = 0.11 $n_{opt}$ and $n_H = -0.32 \times 10^{12}$ cm$^{-2}$ = 0.15 $n_{opt}$, respectively. The Hall effect has single band dependence $n_H = n$ for $n$ approaching charge neutrality.

As in the case of device M2, quantum oscillations in resistance corresponding to twofold degeneracy are observed at low magnetic fields for underdoped $n$ approaching $n_{opt}$ and a metallic region occurs for overdoping between $n_{opt}$ and half-filling at $-n_S/2$. However, the correlated insulator phase is well established at $-n_S/2$ for device D2, irrespective of magnetic field. Superconductivity is absent for $n > -n_S/2$, another notable distinction from the behavior of devices M1 and M2 in [1].

### 3.2.3 Determining $\zeta$

Since the Ginzburg-Landau coherence distance $\xi_{GL}$ exceeds the Moiré superlattice wavelength $\lambda_m = a/[2\sin(\theta/2)]$ by factors ~4 and 3.4, according to results in [1] and [2], respectively, one expects the positional dependence of the pairing interaction to be averaged out. This follows the methodology of [49] for similarly small variations, in which average positions of atomic centers define the interaction distance $\zeta$ and $\xi_{GL}$ generally exceeds interatomic spacings. In applying this method for TBG devices, $\zeta$ is defined as the average of the interlayer separation $z$ weighted by the local electron density $\rho$, both of which vary with Moiré superlattice periodicity [4, 6, 7, 10, 14, 19, 23, 24, 63-69]. Expressed in terms of averages over the Moiré unit cell, one has $\zeta = \langle \rho z \rangle / \langle \rho \rangle$.

Although interlayer separation profiles in TBG structures have yet to be measured directly, thicknesses of ~3.4 to ~4.5 Å at $\theta$ ~ 1° have been reported from AFM measurements (see Fig. 7c of [62]) and periodic superlattice corrugations for $\theta$ near 1° with apparent amplitudes of ~0.3 to over 1 Å, considered to be variously enhanced by electron density modulations, have been observed by STM [4, 63-66]

On uncapped TBG structures, profiles of local density of states (LDOS) by STS and surface topography by STM bear remarkable similarity to one another [4]. As reported in [4] for a sample with $\theta$ = 1.10°, the normalized relative height in their Fig. 1(g) taken at 0.4 V sample bias,[3] and the normalized LDOS in their Fig. 5(c), each exhibit a local maximum at the AA stacking order location in the Moiré superlattice, a minimum at the AB/BA locations and similar saddle points. Writing the interlayer separation as $z = z_{AB} + \Delta z$, where $z_{AB}$ is the minimum of $z$ at AB/BA, $\Delta z \in (0, z_{AA} - z_{AB})$, and $z_{AA}$ is the maximum of $z$ at AA, the findings in [4] indicate that $\rho$ may be taken as proportional to $\Delta z$ such that $\zeta = z_{AB} + \langle (\Delta z)^2 \rangle / \langle \Delta z \rangle$.

---

[3] The ostensibly increased corrugation amplitude near magic angle $\theta$ is strongly dependent upon gate bias, diminishing with increasing bias voltage [65, 66].

Modeling the form of $\Delta z$ as sinusoidal modulations of wavelength $\lambda_m$ with the 6-fold symmetry of a triangular lattice [62, 63] yields the result $\zeta = z_{AA} + f_z(z_{AA} - z_{AB})$ with $f_z = 1/1.8$. Owing to the limited quantitative information available from experiment, theoretical calculations of $z_{AB}$ and $z_{AA}$ [7, 67-69] are used herein. The theoretical $\zeta$ considered are 3.516 Å for $\theta \approx 1°$ from Fig. 4(a) in [67], 3.49 Å for $\theta \approx 1°$ from Fig. 1(d) in [68], 3.504 Å for both $\theta = 0.35°$ and 1.61° in [69], and 3.49 Å for $\theta = 1.08°$ according to the relative height scale of Fig. 1(b) in [7]. Averaging these four theoretical values, $\zeta = 3.50(1)$ Å, where the uncertainty in the least significant digit is the standard deviation. An estimate of the systematic uncertainty can be obtained by modifying the modulations in $\rho$ and $\Delta z$ so as to best replicate the LDOS and normalized height profiles in [4], respectively. By including 5% and 10% additional modulations at periodicity $2\lambda_m$ in $\rho$ and $\Delta z$, respectively, and offsetting $\rho$ by +2.3%, these corrections reduce $f_z$ by 0.04 and $\zeta$ by 0.01 Å for the uncapped TBG device. One can, therefore, infer that the systematic error in $\zeta$ for encapsulated TBG devices appears to be indistinguishable from the theoretical uncertainty. Hence, $\zeta = 3.50(1)$ Å is taken as the best value for device M2 at ambient pressure.

An elastic model is used to deduce the decrease in $\zeta$ under hydrostatic pressure. Since the C-C bonds of the in-plane hexagonal network are generally much stronger than the interlayer van der Waals bonds, it is reasonable to assume a strong anisotropy in elasticity. Assuming an elastic ratio similar to that of graphite [71], the strain induced from hydrostatic compression is strongly dominated by the transverse $z$ component. From the results of calculations presented in [70], a uniaxial compressive strain of magnitude 2.19% is determined for $P = 1.33$ GPa. Scaling from the above result at ambient pressure, the interaction distance for device D2 at 1.33 GPa is $\zeta = 3.42(1)$ Å.

## 4 Discussion

A diversity of theoretical viewpoints has emerged since the discovery of novel superconductivity in gated TBG [5–45], which has been framed as initiating a new era in the superconductivity of materials based on graphite [45]. Superconductivity mediated by electron-phonon interactions, unusually in some instances, and including considerations of non-adiabaticity and Moiré superlattice topology, are found in [5–18]. Superconductivity originating from electronic-based interactions without involving phonons are found in [19–45] and align with the original conclusion for unconventional superconductivity in [1]. This work relates to the latter perspective in that gated TBG superconducting devices have high-$T_C$ behavior and differ markedly from electron-phonon mediated superconductors.

The following explores the theoretical upper limit of $T_C$ and the effects of weak link inhomogeneities for both TBG devices, M2 [1] and D2 [2]. Calculations are then presented on the more uniform D2 device, yielding values for the sheet penetration depth, effective mass, bandwidth and normal-state scattering coefficient. Statistical accuracy of the interlayer Coulomb pairing model is also discussed.

### 4.1 Theoretical Upper Limit of the Transition Temperature

The results in Table 1 follow from the available data on TBG devices and correspond to the restrictions on hole doping of $n_0 > -n_S$ and $n_{opt} < -n_S/2$, which is equivalent to finding $|x_{opt} - x_0| < 2$. The limiting case of half hole filling with unrestricted charge participation is set by $|x_{opt} - x_0| \rightarrow (n_S/2)A = 2$. This defines the theoretical upper limit, expressed as $T_{C0} \leq k_B^{-1} \Lambda (n_S/4)^{1/2} e^2 / \zeta$ from Eq. (2), and corresponds to upper limits of 2.85(1) K for $\theta = 1.05°$ at ambient pressure and 3.53(1) K for $\theta = 1.27°$ at $P = 1.33$ GPa.

For electron doping of the upper flat bands in gated TBG devices, $n > 0$, superconductivity is also developing near half-filling along with the finding of additional insulating phases at one-quarter and three-quarters fillings [2], suggesting a reduced upper limit $|x_{opt} - x_0| \leq 1$ applies to electron doping. Data for optimizing $T_C$ with twist angle and applied pressure are, however, presently incomplete [2].

Wigner crystallization appears to be a viable alternative electronic ordered state for gate-charged TBG at ambient pressure [26] and under compression [27], in competition with superconductivity and inferring a Wigner rather than Mott origin for the insulating states. Close proximity to Wigner crystallization was also noted previously for other high-$T_C$ superconductors [83].

Superconductivity in dual-gated twisted double bilayer graphene devices, which exploit the nearly flat electronic band in graphene, has recently been reported [84, 85]. These new devices are fundamentally unlike single TGB devices, because superconductivity appears for biasing the double layers to differing charge states, breaking interlayer symmetry. Resistance transitions reported thus far lack the signature curvature specific to fluctuation superconductivity, which is evident for single TBG devices (see Fig. 1), owing to an apparent dominance of inhomogeneous broadening; however, upper bounds to the mean field transition temperatures might be assumed from the given 6–7 K estimates of resistance transition midpoints [84, 85].

## 4.2 Uniform Film Versus Josephson Junction Array

In general, vortices form as circulations of currents in a uniform sheet, for which $T_{BKT}$ determines the renormalized sheet magnetic penetration depth, or they form as circulations of currents around loops in arrays of Josephson junctions or weak links, where $T_{BKT}$ sets the renormalized junction coupling energy, expressible in terms of the junction critical current. The universal relation determining $T_{BKT}$ for a uniform two-dimensional superconductor is $T_{BKT} = c^2\hbar^2/16e^2 k_B \Lambda_{BKT}$, where $\Lambda_{BKT}$ is the renormalized sheet penetration depth at $T = T_{BKT}$ [61]. The corresponding relation for two-dimensional arrays of weak links or Josephson junctions is $T_{BKT} = \pi J_{BKT}/2k_B$, where $J_{BKT}$ is the renormalized coupling energy [80, 86, 87].

From magnetic field oscillations in resistance at several carrier densities $n$ bordering "Mott" states, Cao et al. have inferred an inhomogeneous structure in device M1 analogous to Josephson junction arrays [1]. Periodicity of about 9.3 mT is also evident for device M2 near $n = -3.4 \times 10^{12}$ cm$^{-2}$ in Fig. 3b of [1], implying a characteristic area of 0.22 μm$^2$ < $D^2$. Although magnetic field periodicities are virtually absent at $n = n_{opt}$ for superconductivity at highest $T_C$, the current-vs-voltage curve for device M2 displays a steep voltage onset at a critical current of approximately 50 nA [1] (cf. Fig. 1e in [1]), which resembles the abrupt onset of free running phase rotation in a critically biased Josephson junction (an a.c. Josephson effect test could verify this) [88]. Consequently, it is worth examining whether $T_{BKT}$ could be interpreted as the onset of phase incoherence in analogy to the behavior of two-dimensional arrays of weak links [80] and Josephson junctions [86, 87]. Expressing the coupling energy in terms of junction critical current $I_c$ as $J = I_c \hbar/2e$, the universal BKT relation determines the junction critical current as $I_c = 8\varepsilon_c e k_B T_{BKT}/h \approx 46(1)$ nA, given a renormalization constant $\varepsilon_c = 1.8$ as reported in [87] and $T_{BKT} = 0.96(3)$ K. Therefore, junction array behavior at optimal doping appears plausible.

Of particular interest is the proposed expectation [2] that increased applied pressure, with the concomitant increase in the twist angle θ, would improve structural and charge homogeneity due to a reduced TBG Moiré superlattice periodicity. Comparing $\tau_c$ results for devices M2 and D2, superconductivity in device D2 at 1.33 GPa does appear to be more uniform, as is also indicated from the relatively closer proximity of the BKT transition to $T_C$. Comparisons of superconducting properties are made in [2] between a device D1 at ambient pressure with θ = 1.14° and $T_C \approx 0.4$ K (for holes) and device D2 under hydrostatic compression. Superconducting quantum interferences are observed for device D1 (current bias up to about ± 10 mA), whereas none are present for device D2 at 1.33 GPa, and are only weakly present at 2.21 GPa (see Figs. S3C and S3D in [2]). Taking $T_{BKT} \approx 0.27$ K for device D1 at the resistance onset in Fig. 1C in [2] (for holes), the junction array model gives $I_C \approx 13$ nA. In view of this resemblance to the findings from devices in [1], the weak link superconductive behavior of atmospheric pressure TBG devices, as reported in [1] and [2], is evidently a reproducible phenomenon.

## 4.3 Magnetic Penetration Depth and Effective Mass

Accepting the comparatively greater uniformity of device D2 at 1.33 GPa, a renormalized penetration depth $\Lambda_{BKT}$ = (1.96 cm-K) / $T_{BKT}$ = 0.89 ± 0.08 cm and renormalized sheet kinetic inductance $L_{BKT}$ = $2\pi\Lambda_{BKT}/c^2$ = 5.6 ± 0.5 nH follow theoretically from the fitted $T_{BKT}$ = 2.2(2) K.[4] Assuming vortex dielectric constant $\varepsilon_c$ = 1.2 (for small $\tau_c$) [78, 79], the bare sheet penetration depth at $T_{BKT}$ [61] is $\Lambda_s(T_{BKT}) = \Lambda_{BKT}/\varepsilon_c$ = 0.74 ± 0.07 cm. From $T_C^{mf}$ = 2.86(5) K and applying the temperature factor $f_s$ = $1 - (T_{BKT} / T_C^{mf})^4$ = 0.64 ± 0.15 of the Gorter-Casimir two-fluid model (i.e. emulating strong coupling and single gap), one arrives at the zero-temperature value $\Lambda_s(0) = f_s\Lambda_s(T_{BKT})$ = 0.47 ± 0.22 cm. Assuming a thin-film superconductor model with thickness $d = \zeta$ = 3.42 Å, the zero-temperature magnetic penetration depth parallel to the layers is $\lambda_\parallel(0) = (\Lambda_s(0)d/2)^{1/2}$ = 0.9 ± 0.2 μm, thereby providing a Ginsburg-Landau parameter of $\kappa$ = 24 ± 5 for device D2. A simple theoretical estimate from the band structure of magic-angle TBG [89] gives $\Lambda_s(0) = c^2\hbar^2/8\pi e^2\widetilde{D}$ = 0.35 cm, using bare superfluid stiffness $\widetilde{D}$ = 0.31 meV at $n_{opt}$ (geometric mean of x,y components in Fig. 1 in [89]), in reasonable agreement with that calculated here.

The reversed sign of the Hall effect within a region near $-n_S/2$ and enclosing $n = n_{opt}$ is evidence of electron-like carriers coexisting with holes (e.g. potentially forming a superconducting pocket with a small (or zero) gap [22, 28]). Thus, when expressing the penetration depth in reference to the London model, setting $\Lambda_s(0) = m_s c^2/2\pi n_s e^2$, the parameters $n_s$ and $m_s$ in this expression are respectively effective areal carrier density and transport mass of a superconductive state comprising several Fermi pockets. From this perspective, the experimental $\Lambda_s(0)$ is expressible in terms of $n_s/(m_s/m_0)$ = (1.2 ± 0.5) × $10^{12}$ $cm^{-2}$, where $m_0$ is the electron rest mass. Equating $n_s$ in the superconducting condensate to the optimal electron density given by $n_e = n_{opt} + n_S$ = 1.64(2) × $10^{12}$ $cm^{-2}$, one obtains $m_s \approx$ 1.4(7) $m_0$ with a bandwidth of 7(3) meV inferred from $4\pi\hbar^2/Am_s$.

Theory of $m_s$ in flat-band systems is treated in [90].

The transport scattering rate $\tau^{-1} = \alpha k_B T/\hbar$ with numerical coefficient $\alpha$ = 1.1(5) is determined for the normal state near $T \sim T_C$ using the relation $\tau^{-1} = c^2 R_N/2\pi\Lambda_s(0)$ and the normal resistance parameters for device D2 given in Table 2. A coefficient $\alpha$ = 0.2 has been derived from the high-temperature resistance slope and Shubnikov-de Haas data for a magic-angle TBG sample similar to device M2 at optimal gating [73]. The result for device D2 is very similar to normal-state a-b plane scattering rates exhibiting temperature linearity with zero intercept that were previously reported for high-$T_C$ cuprates, e.g., $YBa_2Cu_3O_{7-\delta}$ ($\alpha$ = 1.3(1), $T_C$ = 91 K [91]; $\alpha$ = 2.2(5), $T_C \approx$ 90 K [92]) and $Bi_2Sr_2CaCuO_{8-\delta}$ ($\alpha$ = 2.3(1), $T_C \approx$ 82 K [93]).

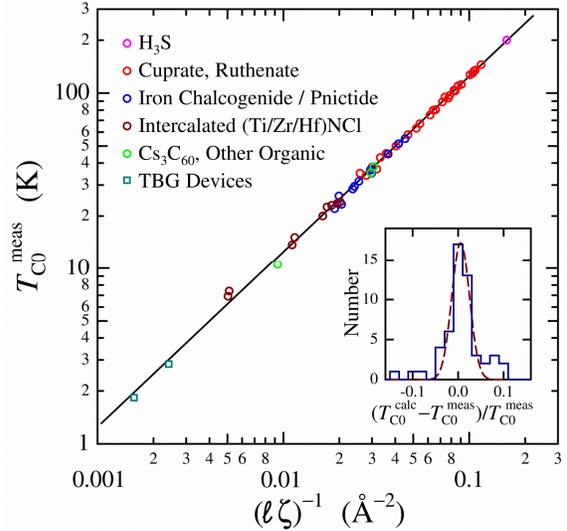

**Fig. 2** $T_{C0}^{meas}$ versus $(\ell\zeta)^{-1}$ for TBG devices M2 and D2 (two square symbols) compared to 51 other optimal high-$T_C$ superconductors: cuprates and ruthenates; iron pnictides and chalcogenides; intercalated group-4-metal nitride-chlorides; and other organics (after [49-56]). The black solid line represents $T_{C0}$ from Eq. (1). Inset shows the distribution (histogram bin width 0.02) of the fractional difference $(T_{C0}^{calc} - T_{C0}^{meas})/T_{C0}^{meas}$ for the 53 optimal high-$T_C$ superconductors, including TBG devices M2 and D2. A fitted normal distribution (dashed curve) is provided for comparison.

---

[4] Together with sheet resistance $R(T_{BKT})$ on the order of 0.5 Ω (roughly estimated by extrapolating and scaling the measured $R(T)/R_{4.5K}$), a relaxation time $L_{BKT}/R(T_{BKT})$ of order 10 ns is predicted for supercurrent flow at $T \approx$ 2.2 K.

### 4.4 Pairing Model Statistical Results

The measured optimal transition temperature $T_{C0}^{meas}$ is plotted against the theoretical factor $(\ell\zeta)^{-1}$ in Fig. 2 for TBG devices M2 and D2 and 51 other optimal high-$T_C$ superconductors: $Im\bar{3}m$ $H_3S$, $Cs_3C_{60}$ (A15 and fcc structural phases), cuprates and ruthenates; iron chalcogenides and pnictides; intercalated group-4-metal nitride-chlorides; and other (BEDT-based) organic superconductors (values taken from [49-56]). The line represents the calculated $T_{C0}^{calc}$ as defined in Eq. (1) and predicts $T_{C0}^{meas}$ with a statistical accuracy of ±1.28 K (root-mean-square deviation between $T_{C0}^{calc}$ and $T_{C0}^{meas}$). For TBG device M2, $T_{C0}^{meas} \equiv T_C^{mf} = 1.83(5)$ K and $T_{C0}^{calc} \equiv T_{C0} = 1.94(4)$ K, and for TBG device D2, $T_{C0}^{meas} \equiv T_C^{mf} = 2.86(5)$ K and $T_{C0}^{calc} \equiv T_{C0} = 3.02(3)$ K.

The accuracy of Eq. (1) is also evaluated by considering the distribution in the fractional difference $(T_{C0}^{calc} - T_{C0}^{meas})/T_{C0}^{meas}$ as is shown in the inset of Fig. 2. Statistics of the distribution are 0.72% mean (indicating that $T_{C0}^{calc}$ is systematically greater than $T_{C0}^{meas}$), 4.3% standard deviation, −0.57 skew, and 2.22 kurtosis. Comparison is made to a normal distribution with fitted 0.55(18)% mean and 1.9(2)% standard deviation (with zero skew and kurtosis by definition) represented by the dashed curve, indicating that about 10% of the data appear as outliers.

## 5 Conclusions

Unconventional superconductivity reported in gate-charged twisted bilayer graphene superlattice devices at ambient [1] and applied hydrostatic pressure [2] is considered to arise from electronic correlations, with characteristics consistent with high-$T_C$ materials. To probe the nature of the superconductive state in TBG, a high-$T_C$ model is proposed where pairing occurs via Coulomb interactions between charges in the two graphene sheets across the separation distance $\zeta$. Compliance with the Coulomb model discussed herein would require the interacting carriers to occupy different graphene sheets. Resistive transitions for devices M2 and D2 were analyzed with generalized Aslamasov-Larkin theory to obtain the experimental $T_C$ as the theoretical mean-field transition temperature $T_C^{mf}$, applied in combination with a generalized Halperin-Nelson interpolation formula to obtain the Berezinskii-Kosterlitz-Thouless temperature $T_{BKT}$. While $T_{BKT} = 0.96(3)$ K obtained for device M2 reflects its weak-link array behavior [1], $T_{BKT} = 2.2(2)$ K found for device D2 is consistent with a uniform thin-film superconductor treatment, thereby yielding $\Lambda_s(0) = 0.47 \pm 0.22$ cm and $\lambda_\parallel(0) = 0.9 \pm 0.2$ μm for sheet and parallel-component penetration depths at zero temperature, respectively.

In analogy with the $Im\bar{3}m$ structural phase of $H_3S$ [50], two identical charge reservoirs, which in the present case are the two graphene layers, each function with coexisting superconductive pairing (type I) and mediating (type II) charges. Identifying the participating charge density as $|n_{opt} - n_0|/2 = 3.0(1) \times 10^{11}$ cm$^{-2}$ from experimental data for fully superconducting device M2 with Moiré pattern and $\zeta = 3.50(1)$ Å defined by the magic angle $\theta = 1.05°$ [1], one finds from Eq. (2) $T_{C0} = 1.94(4)$ K. This result is in remarkable agreement with the experiment; $T_C^{mf} = 1.83(5)$ K and $T_C^{mid} = 1.91(3)$ K, the latter notably larger than the (presumably) similarly estimated ≈1.7 K claimed in [1]. For compressed TBG device D2 with a maximum $T_C^{mf} = 2.86(5)$ K at 1.33 GPa ($\theta = 1.27°$) [2], $\zeta = 3.42(1)$ Å, $|n_{opt} - n_0|/2 = 6.8(1) \times 10^{11}$ cm$^{-2}$ and $T_{C0} = 3.02(3)$ K, again accurately reflecting the experiment.

The unconventional (likely non-phononic) nature of the pairing [1, 2], the required bilayer structure allowing for physically separated charge reservoirs, the stimulation of interlayer excitons [48] and the excellent agreement with Eq. (2), all support the conclusion that the superconductivity in TBG devices is high-$T_C$ in nature. This is a particularly important observation as it establishes TBG as a platform for validating Eq. (1), and understanding the details of the high-$T_C$ superconductive pairing mechanism. With the addition of the TBG M2 and D2 (at 1.33 GPa) devices, validation of the interlayer/interfacial Coulomb interaction pairing model of Eq. (1) is extended to two orders of magnitude in $T_{C0}$ encompassing 53 optimal superconductors from ten disparate superconducting families, predicting transition temperatures with an overall statistical accuracy of ±4% over a range from ~2 to 200 K. Comparisons of $T_{C0}$ with $T_C^{mf}$ from resistance transitions yield accuracies in ($T_{C0}/T_C^{mf} - 1$) of (6.0 ± 3.6)% for device M2 and (5.6 ± 2.2)% for device D2 (at 1.33 GPa).

**Acknowledgments** The authors are grateful for support from the College of William and Mary, New Jersey Institute of Technology and The University of Notre Dame. We also thank Y. Cao for supplemental information.

**Funding** This study was supported by Physikon Research Corporation (Project No. PL-206) and the New Jersey Institute of Technology.

**Compliance with Ethical Standards**

**Conflict of Interest** The authors declare that they have no conflict of interest.

and Jarillo-Herrero P.: arXiv:1901.03710v1 [cond-mat.str-el] (2019)

74. Stepanov N. A. and Skvortsov M. A.: Phys. Rev. B **97** 144517 (2018)

75. Maki K.: Progr. Theor. Phys. **40** 193 (1968)

76. Thompson R. S.: Phys. Rev. B. **1** 327 (1970)

77. Tinkham M.: *Introduction to Superconductivity* 2nd edn. McGraw Hill, New York (1996) chapter 5-5.1

78. Kadin A. M., Epstein K. and Goldman A. M.: Phys. Rev. B **27** 6691 (1983)

79. Fiory A. T., Hebard A. F. and Glaberson W. I.: Phys. Rev. B **28** 5075 (1983)

80. Abraham D. W., Lobb C. J., Tinkham M. and Klapwijk T. M.: Phys. Rev. B **26** 5268(R) (1982)

81. Yoo H., Engelke R., Carr S., Fang S., Zhang K., Cazeaux P., Sung S. H., Hovden R., Tsen A.W., Taniguchi T., Watanabe K., Yi G-C., Kim M., Luskin M., Tadmor E. B., Kaxiras E. and Kim P.: Nature Mater. **18** 448 (2019)

82. Li L., Wang Y., Komiya S., Ono S., Ando Y., Gu G.D. and Ong N.P.: Phys. Rev. B **81** 054510 (2010)

83. Harshman D. R. and Mills A. P. Jr.: Phys. Rev. B 45 10684 (1992)

84. Shen C., Li N., Wang S., Zhao Y., Tang J., Liu J., Tian J., Chu Y., Watanabe K., Taniguchi T., Yang R., Meng Z. Y., Shi D. and Zhang G.: arXiv:1903.06952v1 [cond-mat.supr-con] (2019)

85. Liu X., Hao Z., Khalaf E., Lee J. Y., Watanabe K., Taniguchi T. Vishwanath A. and Kim P.: arXiv:1903.08130v2 [cond-mat.mes-hall] (2019)

86. Leemann Ch., Lerch Ph. and Theron R.: Helv. Phys. Acta **60** 128 (1987)

87. Leemann Ch., Lerch Ph., Racine G-A. and Martinoli P.: Phys. Rev. Lett. **56** 1291 (1986)

88. Heersche H. B., Jarillo-Herrero P., Oostinga J. B., Vandersypen L. M. K. and Morpurgo A. F.: Nature **446** 56 (2007)

89. Hazra T., Verma N. and Randiera M.: arXiv:1811.12428v2 [cond-mat.supr-con] (2018)

90. Törmä P., Liang L. and Peotta S.: Phys. Rev. B **98** 220511(R) (2018)

91. Kamarás K., Herr, S. L., Porter C. D., Tache N., Tanner D. B., Etemad S., Venkatesan T., Chase E., Inam A., Wu X. D., Hedge M. S. and Dutta B.: Phys. Rev. Lett. **64**, 84 (1990).

92. Fiory A. T., Hebard, A. F., Eick R. H., Mankiewich P. M., Howard R. E. and O'Malley M. L.: Phys. Rev. Lett. **65**, 3441 (1990)

93. Romero D. B., Porter C. D., Tanner, D. B., Forro L., Mandrus D., Mihaly L., Carr G. L. and Williams G. P.: Phys. Rev. Lett. **68**, 1590 (1992)
16